\documentclass[aps,prc,12pt,reprint,showkeys,amsmath,amssymb,superscriptaddress]{revtex4-2}
\usepackage[UKenglish]{isodate}
\usepackage[dvipsnames]{xcolor}
\usepackage{soul} 
\usepackage{mathptmx,microtype}
\usepackage{graphicx}
\usepackage{dcolumn}
\usepackage{bm}
\usepackage{enumerate,enumitem,subcaption}
\usepackage[colorlinks=true,       
linkcolor=blue,          
citecolor=blue,        
filecolor=blue,         
urlcolor=blue]{hyperref}
\renewcommand{\frac}{\dfrac}
\DeclareCaptionFormat{custom}{\textbf{#1#2}\textrm{#3}} 
\begin{document}
	\preprint{}
	
	\title{Collective excited states at small amplitude in neutron elastic scattering at low-energies}
	
	
	\author{Do Quang Tam}
	\email[]{doquangtam@hueuni.edu.vn}
	\affiliation{Faculty of Basic Sciences, University of Medicine and Pharmacy, Hue University, Hue City 52000, Vietnam}
	\affiliation{Department of Physics, University of Education, Hue University, Hue City 52000, Vietnam}
	\affiliation{Center for Theoretical and Computational Physics, University of Education, Hue University, Hue City 52000, Vietnam}
	\author{Nguyen Hoang Tung}
	\affiliation{Interdisciplinary Research Center for Industrial Nuclear Energy (IRC-INE), KFUPM, Dhahran, Saudi Arabia}
	\author{Nguyen Hoang Phuc}
	\email{nguyenhoangphuc@hcmut.edu.vn}
	\affiliation{Department of Applied Physics, Faculty of Applied Science, Ho Chi
		Minh City University of Technology (HCMUT), 268 Ly Thuong Kiet
		Street, District 10, Ho Chi Minh City, Vietnam}
	\affiliation{Vietnam National University Ho Chi Minh City, Linh Trung Ward, Ho
		Chi Minh City, Vietnam}
	\author{T. V. Nhan Hao}
	\email[Corresponding author: ]{tvnhao@hueuni.edu.vn}
	\affiliation{Department of Physics, University of Education, Hue University, Hue City 52000, Vietnam}
	\affiliation{Center for Theoretical and Computational Physics, University of Education, Hue University, Hue City 52000, Vietnam}
	
	\date{\today}
	
\begin{abstract}

We investigate the contributions of isoscalar and isovector collective excitations in the neutron elastic scattering of $^{16}$O, $^{40}$Ca, $^{48}$Ca, and $^{208}$Pb nuclei by using a microscopic optical potential (MOP) derived from nuclear structure models based on self-consistent mean-field approaches. Particular attention is given to the role of these collective modes in shaping the imaginary part of the MOP and the resulting angular distributions. Our analysis indicates that both isoscalar and isovector contributions are significant for all considered targets, especially for light and medium targets. Furthermore, the Coulomb interaction is found to play an important role in describing absorption mechanisms and reproducing the experimental angular distributions.

\end{abstract}
	
	
	\maketitle
	
	
\section{Introduction}

At low energies, small amplitude collective motions play a pivotal role in both nuclear structure and nuclear reactions, serving as a bridge between the static properties of nuclei and their dynamic behavior. These motions involve coherent, quantized oscillations of many nucleons around the nuclear ground state and are characterized by their low excitation energy and limited deviation from equilibrium shapes. In the context of nuclear structure, small amplitude collective motions reflect the intrinsic vibrations of the nucleus, such as monopole, octupole, and quadrupole, .etc modes are often described using models like the Random Phase Approximation (RPA), where the excited states are built out of a constructive superposition of particle-hole configurations, which results for example are the low-lying vibrations and the high-lying giant resonances. In nuclear reactions at low energies—such as elastic and inelastic scattering, fusion, and nucleon transfer—small amplitude collective motions can be excited or can couple to the relative motion of the interacting nuclei, significantly influencing the reaction dynamics.

When a nucleon, for example, collides with a nuclei at low energy, the nucleus tends to be excited into collective states, primarily depending on the structure of the nucleus and its deformation. The preferred collective excitation modes in such low-energy nucleon-nucleus interactions are: rotational states for medium to heavy deformed nuclei (especially in rare earth and actinide regions), and vibrational states (spherical or near-spherical nuclei). Different vibrational collective excited states are distinguished by the orbital moment $ L $, spin $ S $ and isotopic spin $ T $ of the excited nucleus. Among them, when the nuclear excited state without isotopic spin change ($ \Delta T = 0 $) is called isoscalar collective excited state, the excited state with isotopic spin change ($ \Delta T = 1 $) is called isovector collective excited state \cite{speth1981giant,sagawa2001giant}.

Each type of mentioned above collective excitation gives us a different vibrational state of the nucleus when it receives the excitation from the incident nucleon. These states are characterized by the nature of the nucleon motion involved: isoscalar excitations involve protons and neutrons moving in phase, while isovector excitations involve protons and neutrons moving out of phase.  However, we do not know whether the nuclei prefer to be excited into isoscalar vibrational collective states and/or isovector vibrational collective states. Understanding how nuclei respond to excitation may give deep insights into: nuclear symmetry properties, constrains nuclear models, inelastic scattering experiments or gamma spectroscopy (can be optimized depending on what kind of excitation is expected), shell structure at far from stability \cite{otsuka2020evolution}.

Recently, modern microscopic approaches, such as Hartree-Fock-Bogoliubov (HFB), and the (Quasiparticle) Random Phase Approximation [(Q)RPA] \cite{dobaczewski1984hartree,fracasso2005fully,avogadro2013role,peru2008role,terasaki2005self,terasaki2010self,colo2013self,niu2016quasiparticle} have demonstrated their predictive power, and computational efficiency, while extending their applicability to exotic nuclei, such as those far from stability or near the drip lines to describe a wide range of nuclear structure observables. Most of these approaches are based on the self-consistent mean-field theories using effective phenomenological $NN$ interactions (which break the explicit link to the bare $NN$ interaction) such as: Gogny interaction (finite-range), and Skyrme interaction (zero-range). The output of these codes can provide the reliable nuclear structure information which can be used to develop the microscopic optical potentials at low-energy.  

In the last decades, the renewed interest in developing microscopic optical potentials to achieve more predictive, systematic, and physically grounded potentials, especially for unstable nuclei \cite{hebborn2023optical}. Most of the microscopic optical potential models (MOP) have been generated from $ab$ $initio$ approaches \cite{quaglioni2008ab,hupin2013ab,hupin2014predictive,nollett2007quantum,waldecker2011microscopic,charity2007dispersive,soma2013ab,hagen2012elastic,navratil2016unified,rotureau2017optical}, nuclear matter approach based on the chiral effective field theory \cite{holt2016microscopic,whitehead2019proton,whitehead2021global,whitehead2020neutron,sammarruca2015toward}, semi-microscopic approaches \cite{bauge1998semimicroscopic,bauge2001lane}, methods based on self-energy \cite{deb2001predicting,Amos2000}, nuclear matter approaches \cite{jeukenne1977optical,barbieri2005nucleon,arellano2002extension,dupuis2006correlations}, and nuclear structure approaches \cite{bernard1979microscopic,mizuyama2012self,mizuyama2014low,blanchon2015microscopic,blanchon2015prospective,blanchon2017asymmetry,nhan2018microscopic,hao2015low,hoang2020effects}. Readers are referred to the most recent comprehensive review on the optical potential for further details \cite{hebborn2023optical}. 
At low energy (below 50 MeV), where the specific nuclear structure effects (such as vibrational collective motions) must be taken into account, the nuclear structure approaches have proved their ability to extract reliable optical potentials directly from the effective phenomenological $NN$ interactions. The energy density functionals built from the $NN$ Skyrme and Gogny effective interactions (even these interactions were not initially designed for scattering purpose) have been successfully applied to reproduce the experimental data on neutron elastic scattering \cite{nhan2018microscopic} by $^{16}$O, proton inelastic scattering \cite{mizuyama2014low} by $^{24}$O, neutron and proton elastic scattering by $^{40}$Ca and $^{48}$Ca and the analyzing power \cite{hoang2020effects, blanchon2017asymmetry}, and neutron elastic scattering \cite{hao2015low} by $^{16}$O and $^{208}$Pb. In the early calculations by V. Bernard $et$ $al.$ \cite{bernard1979microscopic}, the spin-orbit, velocity-dependent, and spin-dependent terms were dropped from the residual interaction within the particle-vibration coupling (PVC) framework due to computational limitations. Also, only isoscalar with $L=0^{+},2^{+},3^{-},4^{+},5^{-}$ and isovector $L=0^{+},1^{-},2^{+}$ states are considered. Subsequently, to reduce the difficulty in the treatment of the continuum of particle-vibration coupling (cPVC) calculations, the two-body spin-dependent terms, spin–orbit interaction, and Coulomb term were omitted in both continuum RPA (cRPA) and cPVC formalisms, as implemented in the work of K. Mizuyama $et$ $al.$ \cite{mizuyama2012self}. 

To date, only two calculations can be considered fully self-consistent: that of G. Blanchon $et$ $al$. \cite{blanchon2017asymmetry}, employing the Gogny interaction, and that of T. V. Nhan Hao $et$ $al.$ \cite{hao2015low,nhan2018microscopic,hoang2020effects}, using the Skyrme interaction. In both studies, the effective interaction is applied consistently throughout the entire framework—from the mean-field to correlations beyond the mean-field. The non-local complex energy-dependent MOP is generated from particle-vibration coupling on top of collective excited stated described by the RPA. These fully self-consistent methods provide a unified description of nuclear structure and reactions within the framework of self-consistent mean-field approaches at low energies. This approach is particularly promising, as it enables a direct comparison between nuclear structure effects and experimental data from nuclear reactions, thereby maximizing the potential to clarify the underlying physical phenomena. In Ref. \cite{hoang2020effects} , the effects of the spin-orbit and velocity-dependent interaction of the effective Skyrme interaction on the angular distributions and analyzing powers of the neutron elastic scattering off $^{16}$O,$^{40}$Ca, $^{48}$Ca, and $^{208}$Pb targets have been studied by using experimental data taken from the National Nuclear Data Center,  Brookhaven National Laboratory Online Data Service \cite{data}. In this work, following the approach of Ref.  \cite{hoang2020effects} we aim to clarify the role of isocalar and isovector collective states during neutron elastic scattering off $^{16}$O,$^{40}$Ca, $^{48}$Ca, and $^{208}$Pb at incident energies below 50 MeV.

\section{Formalism}

\begin{figure*}[!ht]
	\includegraphics[width=1\linewidth]{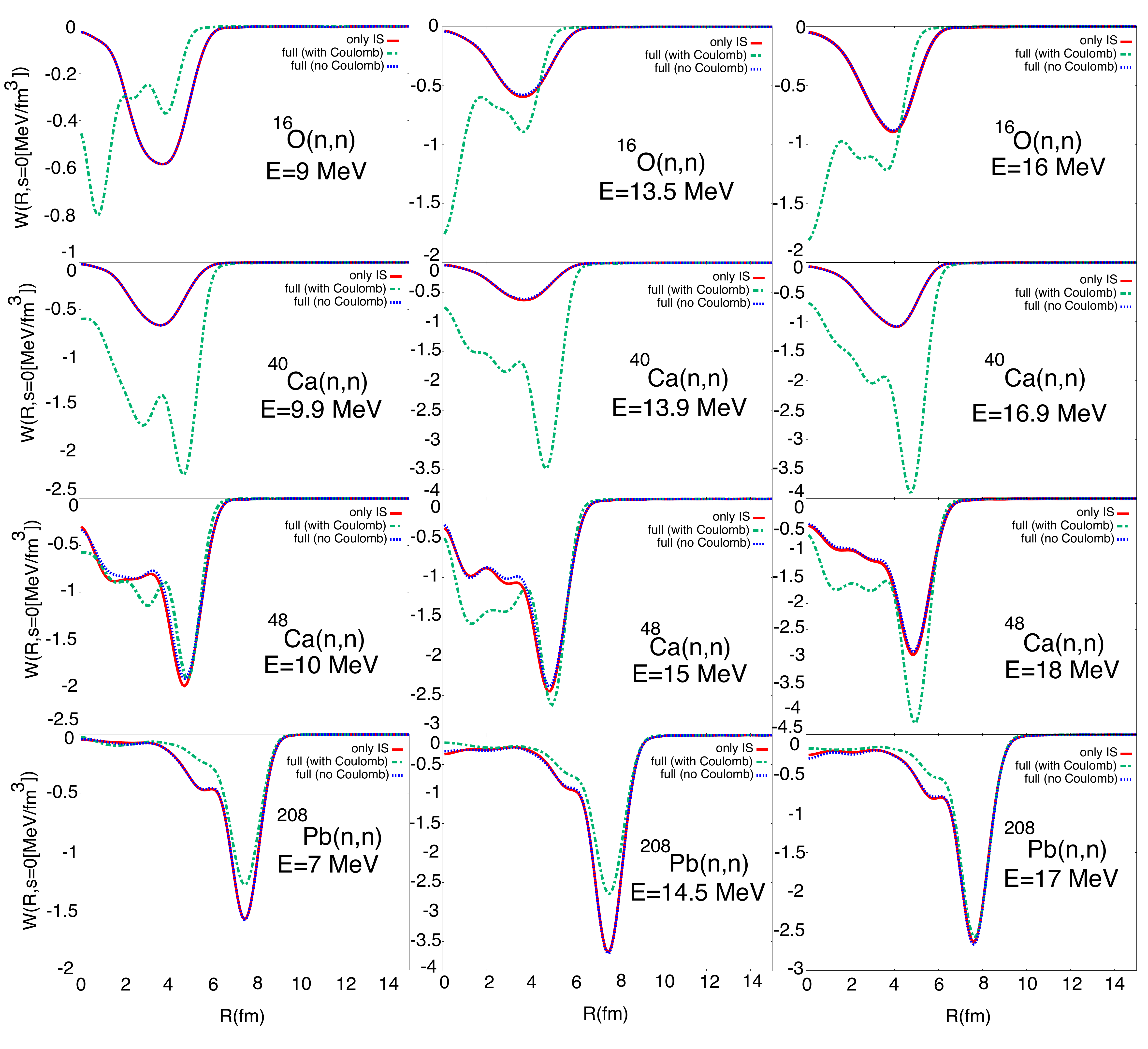}%
	\caption{The calculated W(R, s = 0) for neutron elastic scattering by  $ ^{16} $O, $ ^{40} $Ca, $ ^{48} $Ca, and $ ^{208} $Pb at different incident energies. only isoscalar collectives states-without Coulomb interaction (red solid line), full isoscalar and isovector collectives states-without Coulomb interaction (blue dash-dot line), full isoscalar and isovector collective states-with Coulomb interaction (green dash-dot line).}
	\label{h1}
\end{figure*}

	According to Refs. \cite{hao2015low,nhan2018microscopic,hoang2020effects}, the optical potential is described by:
\begin{equation}
		V_{\mathrm{opt}}=V_{\mathrm{HF}}+\Delta \Sigma(\omega)
		\label{2}
\end{equation}
where
\begin{equation}
		\Delta \Sigma(\omega)=\Sigma(\omega)-\frac{1}{2} \Sigma^{(2)}(\omega).
		\label{3}
\end{equation}
		
In Eqs. \eqref{2} and \eqref{3}, $V_{\mathrm{HF}}$ is a static Skyrme-Hartree-Fock mean field which is real, local, and energy independent proving a major contribution to the real part of the optical potential. The dynamical potential, $\Delta \Sigma(\omega)$, is nonlocal, complex, and energy dependent giving a major contribution to the imaginary part which is responsible for the absorption of the optical potential, and $\omega$ is the nucleon incident energy. The first-order $\Sigma(\omega)$ is the contribution from coupling to the phonon built from the particle-hole excitation. To take into account the issue of the Pauli principle correction, $\Sigma^{(2)}(\omega)$ is the second-order potential (SOP) generated from uncorrelated particle-hole contribution, 
		
	With the imposed spherical symmetry, the partial wave expansion  of $\Sigma(\omega)$ are given by:	
\begin{equation} 
		\begin{aligned}
		\Sigma_{\alpha \beta}^{(l j)}(\omega) \equiv & \left\langle\epsilon_\alpha, l j\|\Sigma(\omega)\| \epsilon_\beta, l j\right\rangle \\
		= & \hat{j}_\alpha^{-1} \hat{j}_\beta^{-1}\left(\sum_{n L, A>F} \frac{\langle\alpha\|V\| A, n L\rangle\langle A, n L\| \| V \| \beta\rangle}{\omega-\epsilon_A-\omega_{n L}+i \eta}\right. \\
		& \left.+\sum_{n L, a<F} \frac{\langle\alpha\|V\| a, n L\rangle\langle a, n L\|V\| \beta\rangle}{\omega-\epsilon_a+\omega_{n L}-i \eta}\right),
	\end{aligned}
\end{equation}
	where $\alpha, \beta$ are generic single-particle states, $a(A)$ denotes the hole (particle) single-particle states, $\epsilon_j$ are the single-particle HF energies, $\omega_{n L}$ are the $n$th phonon energies with multipolarity $L, \hat{j}=(2 j+1)^{1 / 2}$, and the symbol $F$ denotes the Fermi level. The parameter $\eta=1.5$ is introduced to perform the energy averaging on the potential $\Delta \Sigma(\omega)$. The reduced matrix elements $\langle i\|V\| j, n L\rangle$ are the particle-vibration couplings calculated as in Refs. \cite{hao2015low,colo2010effect,cao2014properties}. The effective phenomenological Skyrme interaction has been fully treated and consistently used to describe the residual interaction $V$. To get the nuclear reactions observables, we solve the Schrödinger equation by using the standard DWBA98 code \cite{raynal1998computer} within the nonlocal, complex, and energy-dependent microscopic optical potential $V_{\mathrm{opt}}$ of Eq. \eqref{2}.
	
%
	\section{Result and Discussion}
		\begin{figure*}[!ht]
		\includegraphics[width=1\linewidth]{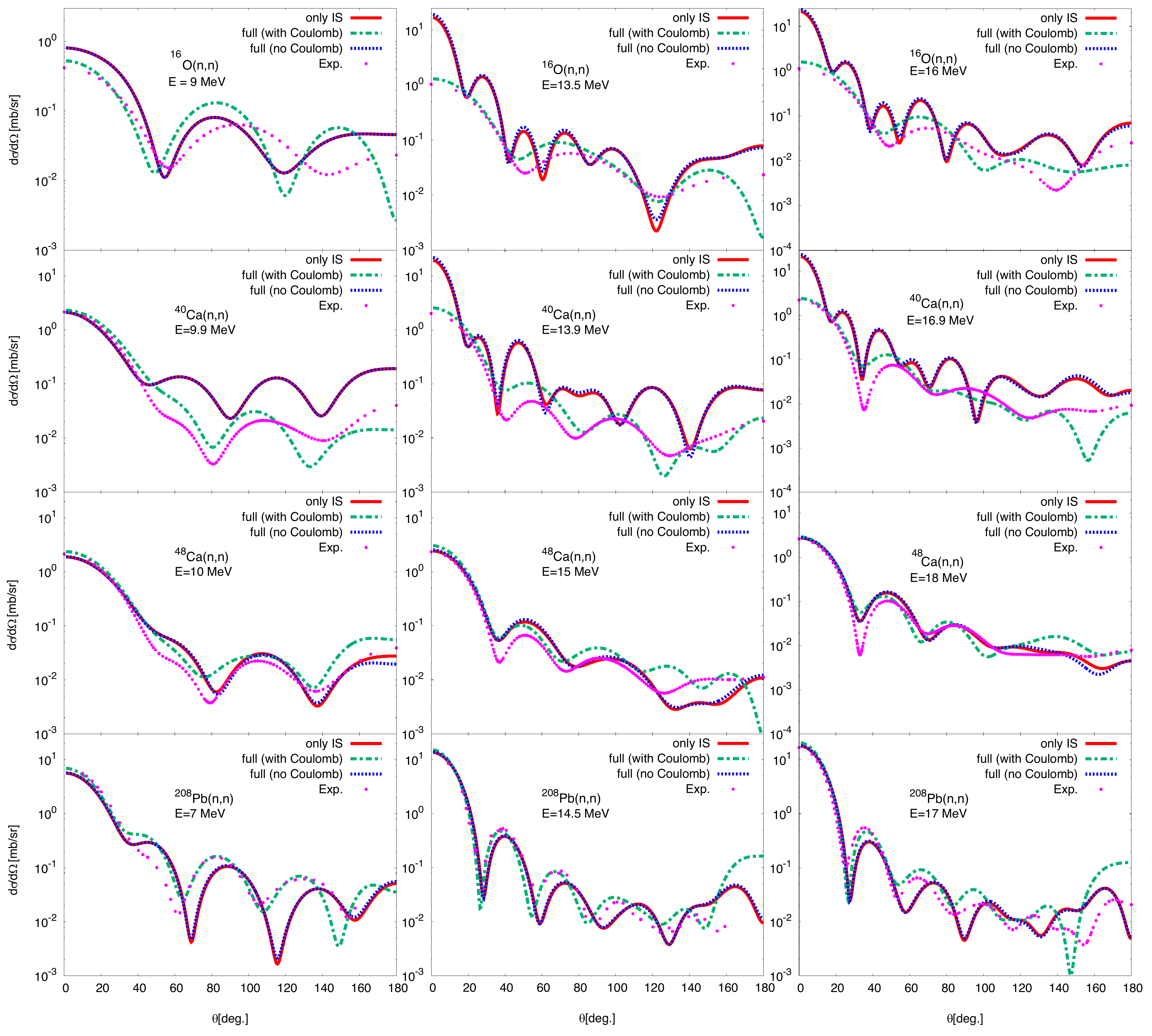}%
		\caption{Angular distributions of neutron elastic scattering by  $ ^{16} $O, $ ^{40} $Ca, $ ^{48} $Ca, and $ ^{208} $Pb at different incident energies below 50 MeV. The interaction SLy5 has been used. Three different calculations are considered: only isoscalar collectives states-without Coulomb interaction (red solid line), full isoscalar and isovector collectives states-without Coulomb interaction (blue dash-dot line), full isoscalar and isovector collective states-with Coulomb interaction (green dash-dot line). The experimental data are taken from Ref. \cite{data}.}
		\label{h2}
	\end{figure*}
The radial Hartree-Fock (HF) equations are solved in coordinate space using a radial mesh size of 0.1 fm, with the radial coordinate extending up to a maximum of 15 fm. The Skyrme-type effective nucleon-nucleon interaction SLy5 \cite{chabanat1998skyrme} is employed consistently across all stages of the calculation, including the mean-field, random phase approximation (RPA), and particle-vibration coupling (PVC) frameworks. The ground state and excited states are constructed on top of the HF solution using fully self-consistent RPA calculations \cite{colo2013self}. We include all natural parity RPA excited states with multipolarity $L=0^{+},2^{+},3^{-},4^{+},5^{-}$ and excitation energies below 50 MeV. To describe doubly magic nuclei, spherical symmetry is imposed throughout the calculation. Continuum states with positive energy are treated by discretizing the spectrum via box boundary conditions. Note that, this continuum has been treated exactly in the work of K. Mizuyama $et$ $al.$ \cite{mizuyama2012self} . Unnatural parity states are excluded from the present analysis. It is important to note that all computational parameters are kept fixed for all nuclei and at all excitation energies considered. 

It is well established that the non-local, energy-dependent imaginary component of the microscopic optical potential (MOP) accounts for the loss of flux from the elastic channel to various non-elastic (reaction) channels, such as inelastic scattering, transfer reactions, and particle emission. To examine the role of isoscalar and isovector collective states on the imaginary part of the optical potential, we present the radial dependence of the diagonal contributions $ W (R, s = 0) $, where $ W (R, s) = \sum_{lj}\frac{2j+1}{4\pi}\textrm{Im}\Delta\sum_{lj}(r, r', \omega) $, where $ R = \frac{1}{2} (r + r') $ corresponds to the radius and shape of $ \textrm{Im}\Delta\Sigma $, and $ s = r - r' $ shows its non-locality.

 In HF-RPA calculations using the effective Skyrme or Gogny interactions, the isoscalar-isovector mixing of vibrational collective states arises due to neutron-proton asymmetry, isospin-dependent parts of the effective interaction, isospin-breaking residual interactions (e.g., spin-isospin, exchange terms, Coulomb effects), configuration mixing, and transition operator structure. As well known, for examples in \cite{sagawa2001giant,peru2005giant,papakonstantinou2011isoscalar}, the Coulomb interaction is the important source of isospin symmetry breaking in the mean-field and RPA residual interaction. We begin by performing three calculations below:
 
 i) full isoscalar and isovector collective states are considered in $\Delta \Sigma(\omega)$ of  Eq. \eqref{2}, the Coulomb interaction is taken into account in the HF mean-field and RPA residual interaction to calculate the  $\Delta \Sigma(\omega)$ of  Eq. \eqref{2}. The obtained results are shown in the green dash-dot line ;
 
 ii)  full isoscalar and isovector collective states are considered in $\Delta \Sigma(\omega)$ of  Eq. \eqref{2}, the Coulomb interaction is turned off in the HF mean-field and RPA residual interaction to calculate the  $\Delta \Sigma(\omega)$ of  Eq. \eqref{2}. The obtained results are shown in the blue dot line ;
 
 iii) only isoscalar collective states are considered in $\Delta \Sigma(\omega)$ of  Eq. \eqref{2}, the Coulomb interaction is turned off in the HF mean-field and RPA residual interaction to calculate the  $\Delta \Sigma(\omega)$ of  Eq. \eqref{2}. The obtained results are shown in the red solid line.
 
 In Fig. \ref{h1}, the three corresponding results for $W(R,s=0)$ are plotted by green dash-dot line for i), blue dot line for ii), and red solid line for iii). The obtained results indicate that the Coulomb interaction plays a dominant role in mixing the isoscalar and isovector contributions for the absorption of MOP. Without Coulomb interaction, the blue line is almost merges with the red line. It also shows that the contributions of isovector collectives states are small in this case. When Coulomb interaction appears, the isovector and isovector-isoscalar mixing contributions change the shape of $W(R,s=0)$ both on the surface and in the interior, especially for light and medium targets. These effects are less pronounced in heavy nuclei compared to those observed in medium-mass and light nuclei. Most of the results (except the heavy nuclei $^{208}$Pb) show that the isovector and isovector-isoscalar mixing contributions increase the absorption in interior of target. At the nuclear surface, the contributions from isovector and isovector–isoscalar mixing can either enhance or reduce the absorption, depending on the target nucleus and the energy of the incident neutrons. Finally, in most cases, isoscalar collective states contribute significantly to the surface absorption. For $^{48}$Ca and $^{208}$Pb, these isoscalar states exhibit a relatively stronger contribution in the nuclear interior compared to those in $^{16}$O and $^{40}$Ca.
 
 The results obtained in this work represent a significant advancement over those reported in the seminal study by V. Bernard $et$ $al.$ \cite{bernard1979microscopic} where they found that the isovector modes contributes very little to $W(R,s=0)$ only for $ ^{208} $Pb. We suppose that it dues to their non-fully self-consistent calculation limited in a reduced model space, including only the isovector states with multipolarities $L=0^{+},1^{-},2^{+}$ and isoscalar $L=0^{+},2^{+},3^{-},4^{+},5^{-}$. Owing to advances in computational capabilities and modern many-body methods, the present work extends those earlier findings by employing a fully self-consistent approach implemented in a significantly broader configuration space ($L=0^{+},2^{+},3^{-},4^{+},5^{-}$  for both isoscalar and isovector states). 
 
 Figure \ref{h2} presents systematic calculations of angular distributions for neutron elastic scattering on  $^{16}$O, $^{40}$Ca, $^{48}$Ca, and $^{208}$Pb at various incident energies, performed for the three cases outlined above. The results indicate that the isoscalar contribution alone is insufficient to reproduce the experimental angular distributions, particularly for light nuclei such as $^{16}$O and medium-mass nuclei like $^{40}$Ca at low energies. In contrast, for the $^{48}$Ca, and $^{208}$Pb targets, the experimental angular distributions are reasonably well reproduced when only the isoscalar contribution is considered, suggesting a dominant role of isoscalar collectives states in these systems.

This work is a further step forward to build a new generation of optical potential which can unify the nuclear structure and nuclear reactions to study the exotic nuclei region. Before approaching the region of unstable nuclei where experimental data are limited, the results presented in this work provide valuable guidance. In its current implementation, the MOP generated from nuclear structure model is primarily applicable to double closed shell target nuclei that are well described within the RPA. Recent studies have performed to address nucleon scattering off nuclei with pairing correlations by employing the Jost function formalism based on the Hartree-Fock-Bogoliubov (HFB) theory \cite{mizuyama2023complex,mizuyama2024eigenphase}. Ongoing developments aim to implement the nuclear structure model through the deformed Quasiparticle Random Phase Approximation (QRPA) to generate the microscopic optical potential to describe the pairing and deformation targets. 

\begin{acknowledgments}
		This research was funded by Hue University under grant no. DHH2025-04-231. T. V. Nhan Hao and Nguyen Hoang Phuc also acknowledge the partial support of Hue University under the Core Research Program, Grant No. NCTB.DHH.2025.17. 
\end{acknowledgments}

	\bibliography{bibliography}
\end{document}